# ASSESSING BIOLOGICAL CONTROL METHOD ONTHE PROGRESSION OF ANAPLASMOSIS DISEASE INDOMINANT CATTLE SPECIES IN THE MATABELELAND NORTH PROVINCE.

Meshach Ndlovu

*Abstract:*This paper presents a compartmental (SI) model for the transmission dynamics of *Anaplasmosis* in resource limited farmer's cattle subjected to a biological control method. The study seeks to evaluate the stability and control of cattle herds dynamics relative to finite agitation. *Anaplasmosis* disease pose a major threat in eradicating cattle population growth in resources limited communities. In gaining the insight of the disease, the following model analysis strategies were used in order to compute simulations, analysis of the model upon varying initial predator population and testing the effects of different predation rate on the disease dynamics. It is essential that the progression of *Anaplasmosis* be stable after the introduction of tick predators into cattle-tick system because that provides the usability of predation as a control measure. After analysing the effect of different prediction rates on the spread of the disease in resource limited communities the study asserted that tick predators like birds and bacteria have been neglected as contributors to natural mechanism of *Anaplasmosis*. Furthermore, the study brought to light that predictors have been neglected as major contributors to natural control mechanism of *Anaplasmosis* in Zimbabwe. Additional numerical simulations showed that predation method can be used in the eradication of *Anaplasmosis* disease thus improving rural livelihood. Investigation of natural tick enemies and predation behaviour can lead to better control of the *Anaplasmosis* disease efficiently and effectively.  Finally, we recommend the necessity for resource limited farmers to capitalise on the use of biological disease control measures.

*Keyword: Anaplasmosis*, biological control, compartmental model, tick eaters.

# I. INTRODUCTION

Anaplasmosis disease is a common cattle eradicator in areas around Matebeleland North and Bulawayo province (Provincial Veterinary Office Bulawayo, 2016). In Zimbabwe, the disease is ranked amongst the top livestock killer diseases, however the country seeks to control this disease through diverse veterinary programs and other policies adopted in the livestock policy (Zimbabwe livestock policy 2012). It is possible for the region to control Anaplasmosis disease if only right measures are implemented. The control measures vary from dipping to vaccination in other cases to treatment (Norval et al. 1984). Anaplasmosis disease in Matebeleland is mainly transmitted by ticks, since ticks are vectors therefore they transmit the disease to cattle. Since the evolution on climate change tick population in Matebeleland North and Bulawayo provinces have also increased as a result of ever increasing global temperature (Stem et al. 1989 and Brown et al. 2012). According to Cunningham (1994), Anaplasmosis disease is transmitted by at least nine different types of tick species . Hence, this paper focuses on tick predation behaviour as a measure for controlling and understanding the progression of Anaplasmosis disease in Matebeleland North and Bulawayo province. Another key factor of consideration is analysing the role played by bio-ecological mechanisms for example predation behaviour in examining the spread of Anaplasmosis disease without drug treatment.

For the past ten years Matebeleland North Province has recorded more than 750 cases of Anaplasmosis with records from the captured records showing that in this region the disease has caused about 60 % mortalities of the reported cases. According to Matebeleland North Database on the disease mortalities section Anaplasmosis disease is ranked number three after Heartwater with 1266 mortalities and Backleg with 657 mortalities. Tick borne diseases accounts to more than 2000 mortality cases from 2011 to 2014. In addition, a survey conducted at areas close to Rhodes Matopos National Parks ranked Anaplasmosis third livestock death contributor with total proportion of 11% (Mlilo et al. 2015).

First, we construct a mathematical model that seeks to explain the current trends of Anaplasmosis disease. The study seeks to explore a cheap bio-ecological method of controlling the Anaplasmosis since most of small holder farmers and commercial farmer cannot manage to vaccinate animals against this disease. Mathematical modeling provides us the tools to measure the progression of Anaplasmosis disease in cattle amongst limited resource livestock ranching (Wangombe, et al., 2009 and Jabbar, et al., 2015).

In Matebeleland North Province cattle farmers have been working tirelessly to eradicate this disease however, the disease has become persistent which has resulted to tick resistance due improper dilution or mix ratio between chemical and water. However, in some regions ticks have evolved to be high tolerant to acaricides (Vidriko et al., 2016, Bandara, and Karunaratne ., 2017). Better modern technologies need to be tried and tested in seeking to end this epidemic. Furthermore, farmers and government came up with different dipping routines but the disease continues to cause death in cattle herds.

## II. Model Construction

In this section, a deterministic Anaplasmosis model with additional predator population unit is formulated and analysed to investigate the role of tick predators in the transmission dynamics of Anaplasmosis in cattle. The model is based on monitoring the dynamics of the disease by constructing a mathematical compartmental model to gain the insight of Anaplasmosis (Diekmann andHeesterbeek , 1999). The populations at any given time (t) are divided into susceptible cattle ($S_B$), infected cattle ($I_B$), susceptible ticks ($S_T$), infected ticks ($I_T$) and predator population unit ($W_p$). Thus, the total cattle population $N_B = S_B + I_B$ and the total tick population is given by $N_T = S_T + I_T$.

**Predators of ticks**
Ticks have a number of natural eaters, these include birds, mammals, arthropods and microbes. Birds that feed on ticks include tickbirds, ox-peckers, chickens and guinea fowls. Arthropods predators include centipedes, ants, spiders and beetles. With such a background we constructed predator population unit and also due to lack of experimental research on the predation rates of tick predators. The study considered a predator population unit to be a finite standard set of all tick eaters which have a significant predation influence.

In the construction of the model the following assumptions were considered.
- The parameter $b_0$ is the influx rate of the cattle population.
- The parameter $b_1$ is the recruitment rate of the ticks.
- The parameter $\mu_B$ is the natural death rate for cattle.
- The parameter $\mu_T$ is the natural death rate for ticks.
- The parameter $\Lambda$ is the disease induced death rate in the infected cattle population.
- A susceptible cattle population $S_B$ can be transferred to the infected sub-population $I_B$ because of an effective transmission due to a bite of an infected tick, at rate $\beta_B$.
- A susceptible tick population $S_T$ can be transferred to the infected subpopulation $I_T$ because of an effective transmission when it bitessome infected cattle, at rate $\beta_T$.

The following Figure 1 shows the developed model.

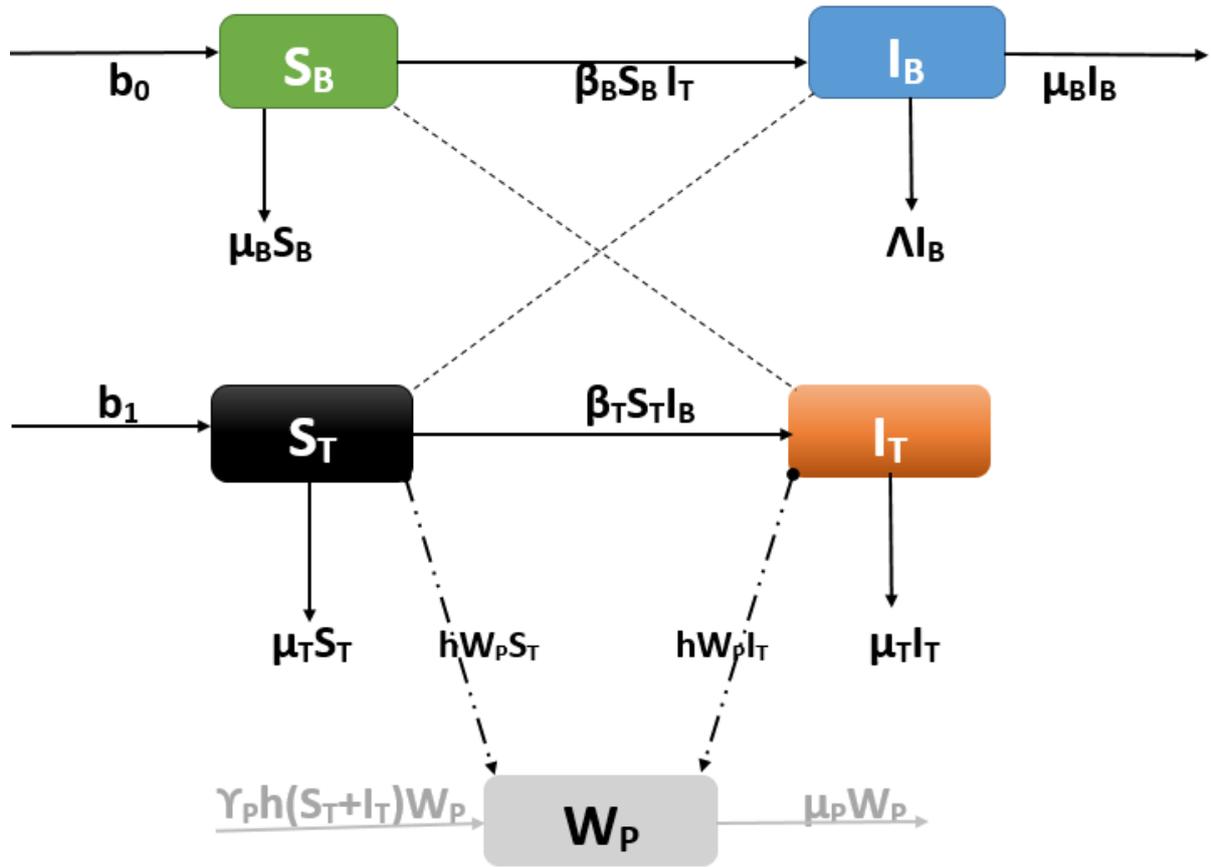

Figure 1: The figure shows a compartment model of Anaplasmosis disease.

Here we explain the variables used in the development of the model in figure 1 where $\Upsilon_p$ is the predator conservation rate, $h$ is the predation rate, $b_0$ is the influx rate of the cattle population, $b_1$ is the recruitment rate of the ticks, $\mu_B$ is the natural death rate of cattle, $\mu_T$ is the natural death rate of ticks, $\Lambda$ is the disease induced death rate in the infected cattle population, $S_B$ susceptible cattle population, $I_B$ infected cattle sub-population, $\beta_B$ effective transmission rate due to a bite of an infected tick. $S_T$ susceptible tick population, $I_T$ infected tick sub-population, $W_p$ is the predator population and the effective transmission rate when a tick bites infected cattle is $\beta_T$.

The broken lines on Fig 1 indicates the interaction between the linked compartments. Thus, on the model equations the terms $hW_p(t)S_T(t)$ and $hW_p(t)I_T(t)$ represents susceptible tick population reduction due to predation and infected tick population reduction due to predation respectively. From the assumptions and model diagram, a deterministic system of non-linear differential equations was obtained.

$$S_B'(t) = b_0 - \mu_B S_B(t) - \beta_B S_B(t) I_T(t)$$
$$I_B'(t) = \beta_B S_B(t) I_T(t) - \Lambda I_B(t) - \mu_B I_B(t)$$
$$S_T'(t) = b_1 - \mu_T S_T(t) - \beta_T S_T(t) I_B(t) - hW_P(t) S_T(t)$$
(1.1)
$$I_T'(t) = \beta_T S_T(t) I_B(t) - \mu_T I_T(t) - hW_P(t) I_T(t)$$
$$W_P'(t) = \gamma_P h[S_T(t) + I_T(t)] W_P(t) - \mu_P W_P(t)$$

With the initial conditions: $S_B(0) \geq 0$, $I_B(0) \geq 0$, $S_T(0) \geq 0$, $I_T(0) \geq 0$, $W_P(0) \geq 0$. Adding the cattle compartments in the model, we have

$$\frac{dN_B(t)}{dt} = b_0 - \Lambda I_B(t) - \mu_B N_B(t)$$
(1.2)

And from the vectors compartments, we obtain

$$\frac{dN_T(t)}{dt} = b_1 - N_T(t)(\mu_T + hW_P(t))$$
(1.3)

Lastly the predator compartment is given by

$$W_P'(t) = \gamma_P h N_T(t) W_P(t) - \mu_P W_P(t) \qquad (1.4)$$

Hence, we consider the invariant space, obtained from equations (1.2), (1.3) and (1.4)

$$\Gamma = \{(S_B, I_B, S_T, I_T, W_P) \in R_5^+, | 0 \leq S_B + I_B \leq \frac{b_0}{\mu_B}, 0 \leq S_T + I_T \leq \frac{b_1}{\mu_T}, 0 \leq W_P\} \qquad (1.5)$$

Therefore, for system (1.1). the region $\Gamma$ is said to be positively invariant.

### III. Model Analysis
*A. Equilibrium points*

The equilibrium states are obtained by setting the right hand side of the system (1.1) of differential equations to zero (Sternberg, S., 2011).

The equilibria of system (1.1) are as follows.
1. Suppose the predator is present, that is $W_P(t) > 0$ for any time(t) such that $t \geq 0$ then there exist a disease free equilibrium point $E_1(S_B, 0, S_T, 0, W_P)$ and a disease equilibrium point $E_2(S_B^*, I_B^*, S_T^*, I_T^*, W_P^*)$ if the reproduction number ($R_1$) is greater than a unit $R_1 > 1$.
2. Suppose the predator is absent, that is $W_P(t) = 0$ for any $t \in [0, \infty)$ then there exist a disease free-equilibrium point $E_3(\frac{b_1}{\mu_B}, 0, \frac{b_2}{\mu_T}, 0, 0)$ and a disease equilibrium point $E_4(S_B^+, I_B^+, S_T^+, I_T^+, W_P^+)$ for reproduction number ($R_1$) greater than a unit $R_1 > 1$.

The introduction of a predator further complicates the system in terms of computing the endemic equilibrium points. In addition, it has been noted from the basic model that the system (1.1) has more than one endemic equilibrium points.

*B. Reproduction Number*
*Lemma 1*

*The disease-free equilibrium point is locally asymptotically stable when $R_1 > 1$ and this means the disease cannot spread. When $R_1 > 1$ then, the disease equilibrium point is said to be unstable and the disease can invade the population.*

The effective reproduction number can be calculated using the next generation matrix method (Driessche and Watmough, 2002). Thus, the effective reproduction number was obtained as:

$$R_1 = \sqrt{\frac{b_0 b_1 \beta_B \beta_T}{\mu_B \mu_T (\mu_B + \Lambda)(\mu_T + hW_P)}} \qquad (1.6)$$

Linear stability of the disease-free equilibrium point is determined by the effective reproductive number. The above reproduction number is given as a function of predator $W_P$ and other parameters. This value of $R_1$ improves the understanding of the stability of *Anaplasmosis* disease in the presence of a predator. Therefore, in order to control *Anaplasmosis* the effective reproduction number $R_1$ should be reduced to a value less than a unit. Thus, predator as a control measure can directly influence the value of $R_1$ since there are inversely related.

### C. Local stability of DFE, $E_1$

The local stability of the disease-free equilibrium point, $E_1$ is described by examining the linearised form of the *system (1.1)* at the disease-free equilibrium point $E_1$. This is done by calculating the Jacobian matrix of the *system (1.1)*. The Jacobian matrix is computed by differentiating each equation in the system with respect to the population variables $S_B$, $I_B$, $S_T$, $I_T$ and $W_P$. The Jacobian matrix at equilibrium point was obtained to be:

Jacobian matrix of the disease-free equilibrium point $E_1$

$$J(E_1) = \begin{pmatrix} -\mu_B & 0 & 0 & -\frac{b_0 \beta_B}{\mu_B} & 0 \\ 0 & -\mu_B-\Lambda & 0 & \frac{b_0 \beta_B}{\mu_B} & 0 \\ 0 & -\frac{b_1 \beta_T}{\mu_T} & -\mu_T-hW_P & 0 & -h\frac{b_1}{\mu_T} \\ 0 & \frac{b_1 \beta_T}{\mu_T} & 0 & -\mu_T-hW_P & 0 \\ 0 & 0 & h\gamma_P W_P & h\gamma_P W_P & h\gamma_P \frac{b_1}{\mu_T}-\mu_P \end{pmatrix} \quad (1.7)$$

The local stability of $E_1$ is determined based on the signs of the eigenvalues of the Jacobian Matrix $J(E_1)$. The disease-free equilibrium point, $E_1$, is said to be locally asymptotically stable if the real parts of the eigenvalues are all negative, otherwise it is said to be unstable. Consider the above Jacobian matrix and let $\phi$ be the eigenvalue. Then we have $|A - I\Phi| = 0$, where A is the matrix $J(E_1)$ and I is a $5 \times 5$ identity matrix. The eigenvalues are:

$$\Phi_1 = -\mu_B < 0 \quad (1.8)$$

$$\Phi_{2,3} = \frac{1}{2\mu_B \mu_T}(\kappa \pm \mu_B \sqrt{(h\ b_1\ \gamma_P)^2 + \mu_T^4 - 4hb_1 h\gamma_P W_P \mu_T - 2hb_1 \gamma_P \mu_P \mu_T + \varpi})$$
(1.9)

Where $\kappa = hb_1 \gamma_P \mu_B - hW_P \mu_B \mu_T - \mu_B \mu_T \mu_P - \mu_B \mu_T^2$ and

$\varpi = 2h(b_1 hW_P \gamma_P \mu_T - W_P \mu_P \mu_T^2 + W_P \mu_T^3) + h\mu_T^2(W_P^2 + 2b_1 \gamma_p) + \mu_P^2 \mu_T^2 - 2\mu_p \mu_T^3$

The real parts of the above eigenvalues $\Phi_{2,3}$ are negative iff: $\kappa < 0$

$$\Phi_{4,5} = -\frac{v}{2} \pm \sqrt{\left(\mu_B \mu_T (\Lambda + hW_P + \mu_B + \mu_T)\right)^2 - 4\tau}$$
(2.0)

Where $v = \Lambda + hW_P + \mu_B + \mu_T$ and

$$\tau = hW_P \mu_B^2 \mu_T^2 (\Lambda + \mu_B) + \mu_B^3 \mu_T^3 \left(\frac{\Lambda}{\mu_B} - \frac{b_0 b_1 \beta_B \beta_T}{\mu_B^2 \mu_T^2} + 1\right)$$

### Lemma 2
*For $R_1 < 1$, the disease equilibrium point is locally asymptotically stable (thus all the real parts of the eigenvalues are all negative). When $R_1 > 1$, the disease-free equilibrium point is unstable (positive or mixture of negative and positive eigenvalues).*

It follows from **Lemma 1** and **Lemma 2**, that the disease-free equilibrium is locally asymptotically stable when $R_1 < 1$. In contrast, if the control measures are not efficient enough such that $R_1 > 1$, then the equilibrium point becomes unstable and the disease outbreak occurs. Therefore, there are three eigenvalues with negative real parts which implies that $R_1 < 1$. For positive eigenvalues $R_1 > 1$ hence the following condition fix them to be negative $hb_1\gamma_P < hW_P\mu_T + \mu_T\mu_P + \mu_T^2$. Thus, we can conclude that the disease-free equilibrium point is locally asymptotically stable when $R_1 < 1$.

## IV. Model Calibration

### A. Study Site
The selection of input values requires livestock diseases trends analysis of the study area so as to estimate parameter values to calibrate the model. Matebeleland North Province comprises of the following districts Hwange, Binga, Bubi, Umguza, Tsholotsho, Lupane, Nkayi and Victoria Falls. The cattle breeds vary within the province from the original Matebeleland Tuli to Nguni, Mashona, and some crosses of all the available breeds in the area. Matebeleland North Province is also a habitant to exotic cattle breeds. Livestock epidemiological data was collected by veterinary staff at dip tanks over a period of more than ten years. Figure 2 shows the distribution of livestock disease rankings from this region.

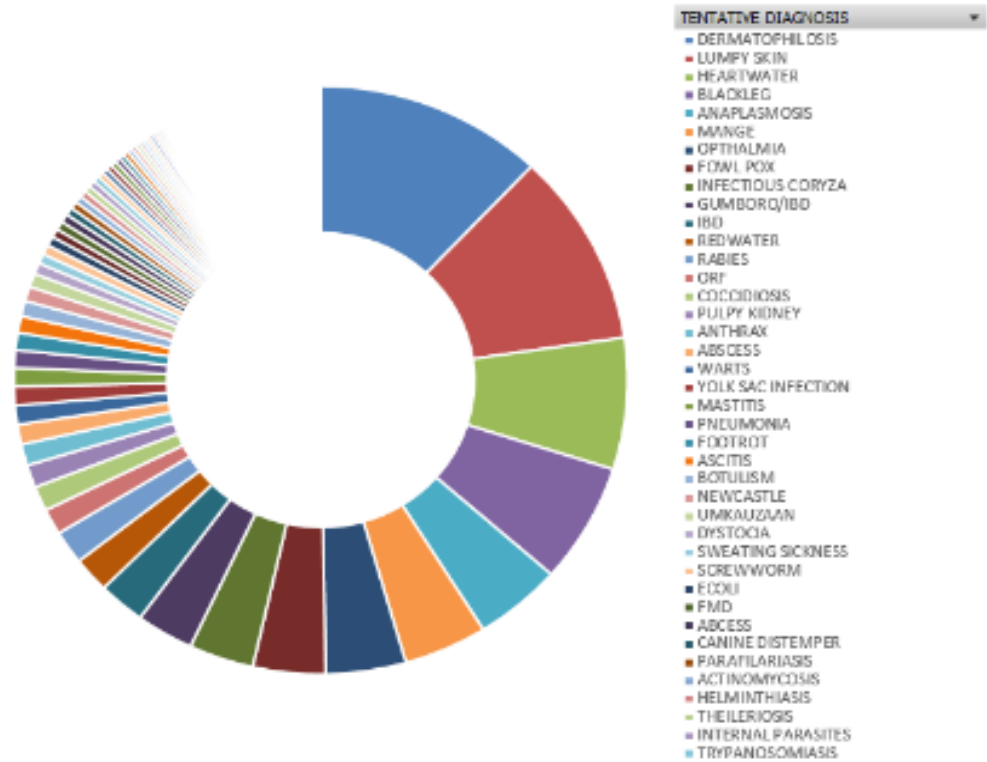

Figure 2: *Ranking of livestock diseases in Mathebeleland Nort Province*

According to *Figure 2 Dermatophilosis* leads the common disease rankings followed by *Lumpy skin*, *Heartwater*, *Blackleg*, *Anaplasmosis*, etc. *Figure 2* also shows epidemiological and disease outbreaks proportions throughout the period. However, the ratio rankings of disease impact in the diagram decreases from the most devastating to the less devastating. The two ends of the doughnut diagram do not meet indicating that there is finite number of livestock diseases some which do not contribute to livestock losses.

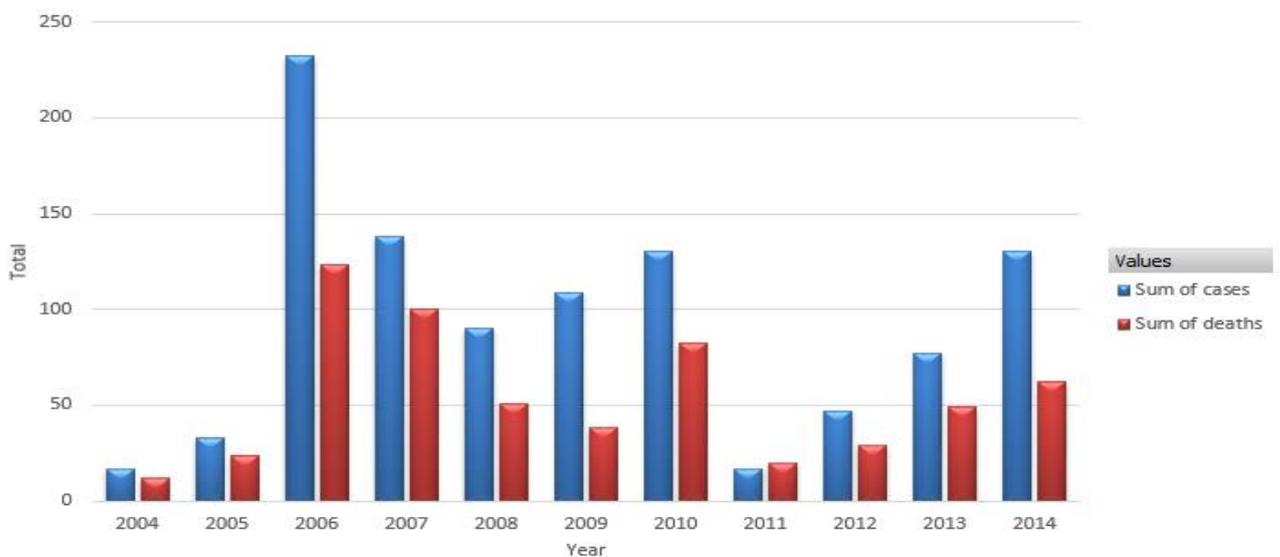

Figure 3: *Anaplasmosis reported cases and deaths in Mathebeleland North Province*

Generally, the reported cases of Anaplasmosis disease are always more than the deaths cases except for year 2011. The year 2006 recorded the highest *Anaplasmosis* incidences and year 2004 recorded the lowest *Anaplasmosis* incidences. Analysing the ten-year period, a periodic trend of the disease is observed. During the year 2011 death cases were more than reported cases, implying that all of the reported cases resulted in death and some deaths can be as a result of rapid impact of the disease leading to deaths and failure by farmers in reporting of such cases. The total cases were 601 and the total deaths were 260 therefore the disease induced death rate ($\Lambda$) can be computed using; $\Lambda = \frac{total\ deaths}{total\ cases} = \frac{260}{601} = 0.4326$.

B.Parameter estimation and simulations

Baseline data was collected from Matebeleland North Province, the secondary data was collected from various publications done by other researchers and some values were assumed. The forecasted values were presented as graphs in order to observe the behaviour of disease after certain parameters in the formulated model were varied. *Table 1* contains the values for parameters used in running simulation of different scenarios.

Table 1: Model parameters and their interpretations

| Parameter | Representing | Value | Source |
|---|---|---|---|
| $\beta_T$ | Cattle to ticks transmission rate | 0.08271 per month | (Rondolph & Rogers, 1997) |
| $\beta_B$ | Ticks to cattle transmission rate | 0.00559 per month | (Medley et al. 1993) |
| $\gamma_P$ | Predator conservation rate | 0.39000 per month | (Lebel et al. 2006) |
| $\Lambda$ | Disease induced death rate | 0.43261 per 1000 | Mat North database |
| $h$ | Predation rate | (0.05- 0.15) | Assumed |
| $\mu_B$ | Cattle death rate | 0.00015 per 1000 | (FAO 2005a) |
| $\mu_T$ | Ticks death rate | 0.35000 per 1000 | (Rondolph & Rogers, 1997) |
| $\mu_P$ | Predator death rate | 0.00500 | Assumed |
| $b_0$ | Cattle recruitment rate | 0.06000 | Assumed |
| $b_1$ | Ticks recruitment rate | 0.40000 | Assumed |

## C. Comparison of different scenarios

The estimated parameters were varied in order to obtain different models simulations. It is critical to vary parametric values so as to clearly understand the behaviour of the model under different scenarios.

Table 2: Different scenarios considered when running the simulations

| Scenario | Parameter | Value used |
|---|---|---|
| 1 | $W_P$ | 0 |
| 1 | $h$ | 0 |
| 2 | $W_P$ | 2 |
| 3 | $W_P$ | 10 |
| 4 | $h$ | 0.05 |
| 4 | $h$ | 0.1 |
| 4 | $h$ | 0.15 |

We consider cattle populations on a single farm with the initial cattle population and tick population estimated values of $S_B = 140$, $I_B = 50$, $S_T = 156$, $I_T = 100$ and for the other initial conditions we consider the parameter values as given in Table 1.2.

# V. Results

To present the progression of *Anaplasmosis* graphical and numerical summaries for model analysis were plotted and interpreted.

**Scenario One**

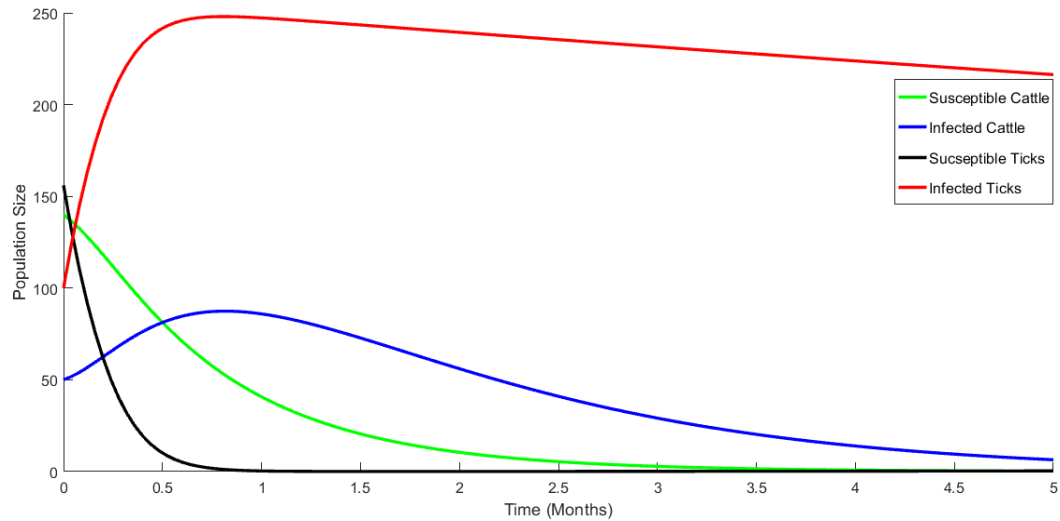

Figure 4: *Numerical simulation for scenario one which is the model without a predator.*

*Figure 4* shows the simulation curves of basic model (model without treatment) hence, the simulations graphs were considered as point of reference with other results from model modifications. The susceptible cattle population reduces exponentially until it reaches the lower bound which in this case is zero and accounting for such an exponential decay to be as a result of low cattle influx as compared to the out-flux. Once again the susceptible tick population curve also exhibits rapid exponential reduction as a result of low recruitment rates of tick population as compared to low removal rates. Infected tick population graph experiences a rapid exponential growth before reaching the peak and then, after the turning point the graph slowly decreases. Infected cattle population graph initially grows steadily to the maxima then from the maxima the population reduces steadily approaching zero. Without a control measure it is expected that the disease persists, what is then left it to observe the response of the disease dynamics with a predator population

**Scenario Two**

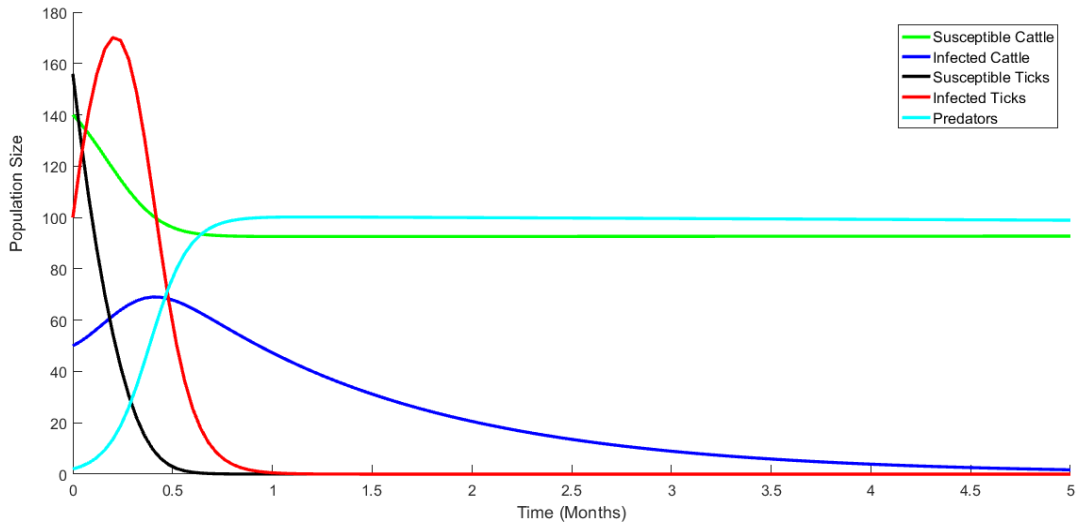

Figure 5: *Numerical simulation for scenario two which is the model with two initial predators.*

The changes observed by simple introduction of two predator population units into the system are depicted in *Figure 5*. Susceptible cattle population curve collapses to a lower limit bound greater than zero as compared to the same population curve in *Figure 4*, with lower bound changes from zero to about 112. Considering susceptible tick population curve, the population reduces rapidly and faster as compared to the same curve on *Figure 4*. However, the infected tick curve shows a rapid growth of the population to reach the maxima and then rapidly decreases matching that of susceptible tick population. In this case, the tick population decrease to zero whereas the cattle population approaches a constant. In addition, the peak population size in *Figure 4* for infected ticks was approximately 250 whereas in *Figure 5* it was around 170. Without a predator the infected tick population hardly reduces however, the predator effect is evidently shown on *Figure 5* to speedily reduce the tick population.

The infected cattle population curve is belly shaped and skewed to the left which implies that, an increase of new cattle infection occurs soon after the disease invasion and then the disease tails off slowly. Scenario two simulations indicated a rapid suppression of tick population with just introduction of two predator population units. One of the assumption made on model development was that the recruitment is strictly not dependent on birth rate thus the populations can growth rapidly within a short period of three to four weeks. Therefore, the suggestion of such a speedy growth in predator population might be as a result of immigration due to the availability of food. For the first few weeks, the infected cattle population survive even in low predation rate equivalent to the disease gestation period. The use of predators to control the disease do not directly affect the infected cattle population as it decreases slowly after predator have reached peak. It has been shown that predation method can be useful in controlling *Anaplasmosis*, now follows the investigation of progression of the disease subjected tonumerous predator population.

**Scenario Three**

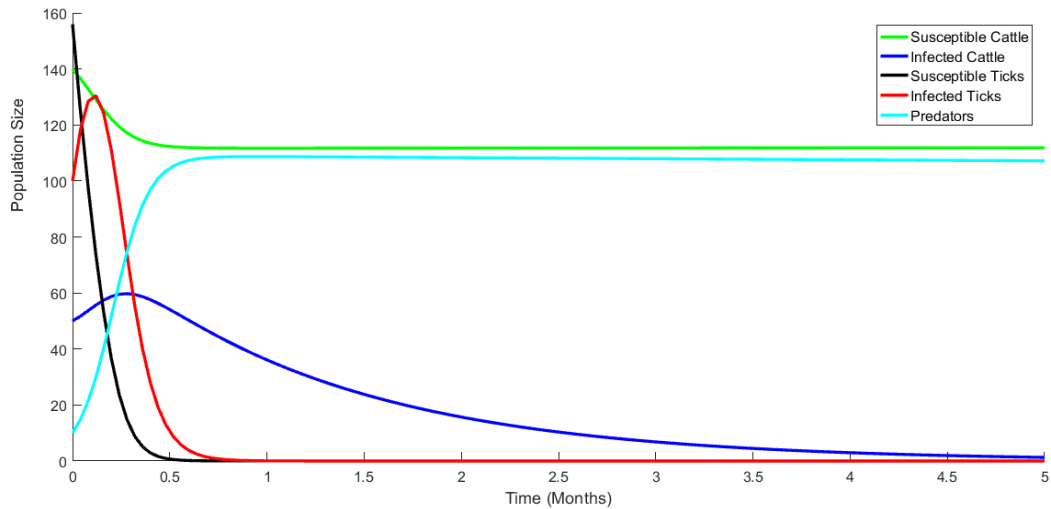

*Figure 6: Numerical simulation for scenario three which is the model with ten initial predators.*

After increasing initial predator population, the observed response simulations were computed shown on *Figure 6*. The behaviour of the *system (1.1)* on scenario three exhibits slight differences with meaningful implications from the previous *Figure 5*. Carrying out further comparisons on population dynamics using figure 2 and *Figure 4* we noticed that, the peak for infected tick population decreases from about 170 to 125. The maxima for infected cattle drops from approximately 70 to 60 and the lower bound for the susceptible reduces from about 115 to 90. Finally, the susceptible tick population reduces to zero faster than before. An observation on *Anaplasmosis* progression is that predators can reduce the number of infectious population thus reducing the disease and also (Richards *et.al. 2006)* assert to the same fact.

The predator population grows following the Lokta-Volterra logistic growth rules, until it reaches the maximum carrying capacity then it remains constant for some time. Analysing the total cattle population, it remains constant at the beginning because the increase of infected cattle seems to be proportional to the decrease of the susceptible cattle population. After that the susceptible cattle remains fixed as the infected cattle population slowly approaches zero which implies predators can help in the eradication of the disease. Soon after introducing ten predator population units there is quick removal of tick which in-turn implies that there are less losses of cattle.

**Scenario Four**

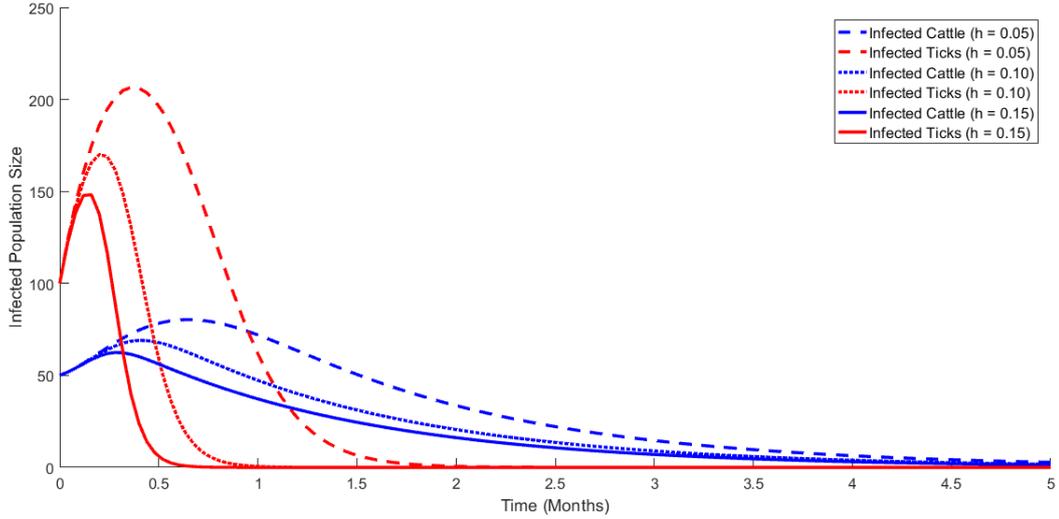

*Figure 5: Numerical simulation for scenario four which is the model of infected population with varying predation rates.*

The other parameter of interest is the predation rate (*h*), by varying the predation rate (*h*) from 0.05 to 0.15 the infectious population size tremendously reduces as the value of (*h*) increased. On a special note we noticed a key result that the difference between two preceding infected population at (∆(*h*)) becomes smaller as predation rate (*h*) say ∆(*h*) → *δ* where *δ* is a value such that the simulation graph of two preceding infected populations are merged together. Thus, explaining the net change to be ∆(*h*) ≈ 0. Eventually the infected population will converge to a certain value at a certain point which is the equilibrium point. Varying the predation rate is a method to investigate another aspect of reducing tick population, thus the research results obtained from scenario four indicates that predation rate (*h*) strongly impact on the infected population by reducing the number of the infectious population. Hence, by considering the infected cattle population, the time to deaths for infected animals remains constant, this is evident since all the three simulations start and end at the same points. We conclude that the predation rate has the same effect as the predator population size on controlling the impact of *Anaplasmosis* disease.

## VI. Discussions and recommendations

In exploring knowledge gained from the simulation outputs we compared and analysed the feasibility of the predation method versus other control methods.

*Sensitivity Analysis:* The reproduction number for the basic model calculated using the values on *Table (1.1)* and the initial conditions $h = 0$ $R_0$ was calculated to be $R_0 = 0.29797 < 1$; the disease equilibrium point of basic model (without predators) implies that the disease can die out natural on its own. For $R_1 = 0.77208 < 1$ the disease equilibrium point of the model with a small number of predator implies that the disease can die out.

When $R_1 = 1,62033 > 1$ then, the disease equilibrium point of the model with large predator population is globally asymptotically stable. This implies that, the disease persists in the presence of large predator populations. This might be true as some authors highlighted that predators can act as carriers of the *Anaplasmosis* (Gilbert *et al*., 2004). Furthermore,

Sekercioglu, (1996) suggested that there is no solid evidence that birds can control tick population as they themselves can be host for ticks. In addition, *Anaplasmosis* disease can be eliminated if predator population and predation rate is large enough such that the function $R_1(W_P; h)$ dependent on $W_P$ and $h$ to less than one $R_1(W_P, h) < 1$.

*Predation rate:* The predation rate is also a significant player in the reduction of *Anaplasmosis* disease. An increase in the predation rate effects a drop in the infected population size. However, less work has been done on the estimation of this parameter in real world scenarios. The simulations of the model showed that predators of ticks in rangelands play a very important role in the spread of *Anaplasmosis* disease. Predators have for years been neglected as amongst the factors responsible for the controlling of *Anaplasmosis*. In addition, our model suggests that increasing the predators and predation rate can be influential in controlling the spread of *Anaplasmosis*. Hence, resource limited farmers are advised to capitalise on the use on tick predation method.

### VII. Discussions on remaining questions and future studies

An investigation on the natural tick enemies behaviour can lead to better efficient and effective control of the *Anaplasmosis* disease. One of the key question is how to increase the predator population so as to increase predation rate? In seeking to improve the predator situation on rural communities there is great need for proper patience, planning and great deal of research. Another important factor is the education of resource limited farmers on environmental conservation so that their communities prove to be conducive habitation of tick predators. To effectively implement predation methods as a control measure for tick borne disease there is need for experimental research to back up these mathematical models by comparing bio-control measures with vaccination and treatment methods. However, vaccines are expensive most of the resource limited farmers in Matebeleland North Province cannot afford. If prevention is better than cure, then further studies should be conducted in order to analyse the dietary habits of the biological control. The disadvantage of using acaricides for controlling tick population is because of their harmfulness to the environment, animals and humans.

Most of the times when we analyse treatment and vaccination models the main focus is on the dynamics of the infected cattle population whereas, for models with predators as control measure the focus is on conservancy and preservation of the susceptible cattle population. It has been shown that in the presence of predators many cattle can survive the impact of *Anaplasmosis* disease. The fact that predators can act as carriers of disease, thus too large numbers of predators can force the disease to persist. Hence, there is a greater need to study the optimal predator population sizes required to control the disease.

### REFERENCES

Bandara, J and Karunaratne P., (2017). Mechanisms of acaricide resistance in the cattle tick *Rhipicephalus (Boophilus)microplus* in Sri Lanka. Pestic Biochem Physiol. 2017 Jun; 139: 68–72.

Brown, D., Chanakira, R.R., Chatiza, K.,Dhliwayo, M., Dodman, D., Masiiwa, M., Muchadenyika, D., Mugabe, P., Zvigadza, S., (2012). Climate change impacts, vulnerability and adaptation in Zimbabwe IIED Climate Change Working Paper No. 3, December 2012.

Castillo-Chavez, C., Feng, Z., Huang, W., (2002). On the computation of $R_0$ and its rolein global stability, in: C. Castillo-Chavez, withS. Blower, P. van den Driessche, D. Kirschner, A. Yakubu (Eds.), Mathematical Approaches for Emerging and Reemerging Infectious Diseases: An Introduction, Springer, 2002, p. 229.

Comprehensive Agriculture Policy Framework, 2012. Livestock Sector Policy Issues and Statements, improve animal health and welfare.
Retrieved fromhttp://extwprlegs1.fao.org/docs/pdf/zim149663.pdf

Cunningham, M.P. (1981). Biological control of ticks with particular reference to *Rhipicephalus appendiculatus*. In:Advances in the Control of Theileriosis.

Diekmann O, Heesterbeek J.A.P, (1999) Mathematical Epidemiology of Infectious Diseases: Model Building, Analyis and Interpretation, Wiley, New York.

Food and Agriculture Organization FAO.,(2005a). Livestock Sector Brief, Zimbabwe.
Retrieved from
http://www.fao.org/ag/againfo/resources/en/publications/sector_briefs/lsb_ZWE.pdf

Gilbert, L., Jones, L.D., Laurenson, M.K., Gould, E.A, Reid, H.W., and Hudson,P.J. (2004). Ticks need not bite their red grouse hosts to infect them with Loupingill virus. Biology Letters. 271: 202-2051q Recruitment

Jabbar A, Abbas T, Sandhu Z-D, Saddiqi HA, Qamar MF, Gasser RB (2015). Tick-borne diseases of bovines in Pakistan: major scope for future research and improved control. Parasites & Vectors. 2015;8:283. doi:10.1186/s13071-015-0894-2.

Lebel, C. Olsson, P., L. H. Gunderson, S. R. Carpenter, P. Ryan, L. Folke, and C. S. Holling (2006). Shooting the rapids: navigating transitions to adaptive governance of social-ecological systems. Ecology and Society 11(1): 18. Retrieved from http://www.ecologyandsociety.org/vol11/iss1/art18/

Medley, G.F., Perry, B.D., Young, A.S., (1993). Preliminary analysis of transmission dynamics of Theileria parva in eastern Africa. Parasitology 106, 251-264.

Mlilo D, Mhlanga M, Mwembe R, Sisito G, Moyo B, Sibanda B (2015) The epidemiology of Malignant catarrhal fever (MCF) and contribution to cattle losses in farms around Rhodes Matopos National Park, Zimbabwe. Trop Anim Health Prod 47(5):989–994.

Mwambi, H. (2002) Ticks and tick borne diseases in Africa: a disease transmission model. IMA J. Math. Appl. Med. Biol., 19, 275-292.

Norval R.A.I., Barrett J.C., Perry B.D. and Mukhebi A.W. (1992b). Economics, epidemiology and ecology: A multi-disciplinaryapproach to the planning and appraisal of tick and tick-borne disease control in southern Africa.

Provincial Veterinary Office Bulawayo Livestock, (2016). Database unpublished raw data.

Randolph, S.E., Rogers, D.J., 1997. A generic population model for the African tick *Rhipicephalus appendiculatus*. Parasitology 115, 275-279.

Richard S. Ostfeld, Amber P, Victoria L. Hornbostel, Michael A. Benjamin, and Felicia Keesing., (2006) Controlling Ticks and Tick-borne Zoonoses with Biologicaland Chemical Agents. Retrieved from
http://www.caryinstitute.org/sites/default/files/public/reprints/Ostfeld et.al., 2006BioSci 56

Sekercioglu, (1996) No solid evidence that guinea fowl control tick populations. Retrieved from http://www.examiner.com/article/no-solid evidence-that-guineafowl-control-tick-populations

Shlomo Sternberg May (2011). Dynamical Systems pp 221 Retrieved from
http://www.math.harvard.edu/library/sternberg/text/book.pdf

Stem, E., Mertz, G.A., Stryker, J.D. and Huppi, M., (1989). Changing animal disease patterns induced by the greenhouse effect. The Potential Effects of Global Change on the United States: Appendix C - Agriculture, Volume 2. Eds: Smith, J. and Tirpack, D.A., US Environmental protection agency, Washington, D.C. pp. 11-38.

Van den Driessche. P and Watmough. J, (2002). Reproduction numbers and subthreshold endemic equilibria for compartmental models of disease transmission.

Vudriko, P., Okwee-Acai, J., Tayebwa, D. S., Byaruhanga, J., Kakooza, S., Wampande, E., Omara, R., Muhindo, J. B.,Tweyongyere, R., Owiny, D. O., Hatta, T., Tsuji, N., Umemiya-Shirafuji, R., Xuan X., Kanameda, M., Fujisaki, K. andSuzuki, H. (2016). Emergence of multi-acaricide resistant Rhipicephalus ticks and its implication on chemical tickcontrol in Uganda. Parasites & vectors, 9, 4. doi:10.1186/s13071-015-1278-3

Wangombe, A., Andersson, M. and Britton, T. (2009). A stochastic epidemicmodel for tick borne diseases: initial stages of an outbreak and endemic levels(submitted).